# Nanoscale displacement sensing based on nonlinear frequency mixing in quantum cascade lasers

F. P. Mezzapesa, L. L. Columbo*, G. De Risi, M. Brambilla, M. Dabbicco, V. Spagnolo and G. Scamarcio

*Abstract*— We demonstrate a sensor scheme for nanoscale target displacement that relies on a single Quantum Cascade Laser (QCL) subject to optical feedback. The system combines the inherent sensitivity of QCLs to optical re-injection and their ultra-stability in the strong feedback regime where nonlinear frequency mixing phenomena are enhanced. An experimental proof of principle in the micrometer wavelength scale is provided. We perform real-time measurements of displacement with λ/100 resolution by inserting a fast-shifting reference etalon in the external cavity. The resulting signal dynamics at the QCL terminals shows a stroboscopic-like effect that relates the sensor resolution with the reference etalon speed. Intrinsic limits to the measurement algorithm and to the reference speed are discussed, disclosing that nanoscale ranges are attainable.

## I. INTRODUCTION

THE development of compact, monolithic and reliable electrically driven laser sources in the mid-infrared to terahertz (THz) range of the electromagnetic spectrum represented by the Quantum Cascade Lasers (QCLs) has triggered in the last years the emergence of promising technologies for applications in a variety of fields ranging from industrial process and quality control, to imaging and security, medical diagnosis, spectroscopy, communications technology, space science [1].

Notably, continuous wave emission (CW) in QCLs is intrinsically ultra-stable against strong optical injection [2], tolerating feedback levels which typically cause dynamical instabilities in bipolar semiconductor lasers, such as mode-hopping, intensity pulsation or coherence collapse [3]. As we recently demonstrated [2], this property directly follows from i) the absence of relaxation oscillations (class-A laser) owing to the ultrafast intersubband relaxation time in these unipolar devices; ii) the small linewidth enhancement factor.

This has led to the realization of compact, real-time, easy-to implement sensors for applications in coherent imaging [4], motion tracking [5], material processing [6] based on laser self-mixing (SM) in QCLs. In the single arm SM interferometric configuration [3] the coherent superposition between the laser intracavity field and the radiation back reflected form an external target allows to directly correlate changes in the object target "state", e.g. in terms of position relative to the QCL exit facet, with voltage modulations across the QCL terminals.

Recently, ever-growing real world applications in nanofabrication have increased the demand for nanometer resolution in laser position and displacement sensors. Several systems based on the optical feedback interferometry, employing standard algorithms to off-line enhancing the close-loop performances, have been proposed [7-10]. The main drawback is that at least one fringe is needed to implement the signal analysis process for phase retrieval. Prototype systems featuring on-line nanometric resolution have been lately demonstrated in differential feedback interferometry referenced to a twin diode laser [11, 12]. An open question is whether nanoscale resolution can in principle be achieved in laser feedback-based sensing systems working at longer wavelengths, including those using QCLs.

In this paper we demonstrate a sensing technique to determine displacement with subwavelength resolution by exploiting the nonlinear frequency mixing that can be achieved in a QCLs subject to optical feedback, precisely due to its enhanced stability in the strong feedback regime [2]. The sensor scheme is sketched as in Fig. 1, where a reference target (beam splitter), denoted in the following as RT, is inserted in the external cavity formed by the QCL and the object target, denoted as OT. The multiple beams reinjected into the QCL coherently interfere with the intracavity field, yielding to frequency mixing phenomena that brings information on both the RT and the OT motion. In particular, we show that signal sampling of fringes from OT can be performed at shorter intervals when RT speed increases. Such denser sampling allows improving the resolution of the OT displacement measure well below the half-wavelength limit, typical of self-mixing-interferometers. We show that the nanometer scale can be approached without resorting on post-

F. P. Mezzapesa, L. L. Columbo, G. De Risi, B. Brambilla, M. Dabbicco, V. Spagnolo, and G. Scamarcio are with the Dipartimento di Fisica, Università degli Studi e Politecnico di Bari, and CNR - Istituto di Fotonica e Nanotecnologie UOS Bari, via Amendola 173, I-70126 Bari, Italy
(e-mail:francesco.mezzapesa@uniba.it,lorenzo.columbo@uniba.it, giuseppe.derisi@gmail.com,massimo.brambilla@uniba.it,maurizio.dabbicco @uniba.it vincenzo.spagnolo@uniba.it, gaetano.scamarcio@uniba.it).

processing via signal analysis. Differently from the modulation scheme reported in [13-16] the proposed sampling reference method allows to improve the resolution of any feedback interferometric scheme, widening the application range of QCL-based sensors to include depth-resolved THz imaging with sub-wavelength accuracy and nano-step height profiling in materials opaque to visible light.

A careful theoretical analysis of the QCL subject to optical feedback from multiple cavities is presented in sec.II, while in sec.III a the displacement retrieval algorithm implemented to reach a sensitivity of λ/100 is described and discussed. A valuable proof-of-principle experiment of high-resolution displacement sensing based on a commercial MIR-QCL is provided in sec.IV. Finally, sec.V is devoted to draw our conclusions and future developments.

## II. THE THEORETICAL ANALYSIS

The analysis of the Lang-Kobayashi (LK) model at steady state, in presence of two partially reflecting targets (see Fig.1), was introduced in [17]. We assume in the following that the displacement of the OT is the physical quantity we want to measure, and the motion of the RT is fully determined and known with arbitrary accuracy. We recall that, when the OT and the RT move with constant speeds $v_1$ and $v_2 < v_1$ respectively, the laser frequencies are given by the transcendental equation:

$$\omega_F = \omega_0 - \frac{k_1}{\tau_c}\left[\alpha\cos(\omega_F\tau_1) + \sin(\omega_F\tau_1)\right]$$
$$- \frac{k_2}{\tau_c}\left[\alpha\cos(\omega_F\tau_2) + \sin(\omega_F\tau_2)\right] \quad (1)$$

where, $\omega_0$ is the free running QCL frequency, $\tau_c$ is the field round trip time in the QCL cavity, $k_i$ are the feedback coefficients and $\alpha$ is the linewidth enhancement factor. The delays $\tau_i$ change in time due to the target motions so that $\tau_i = 2L_i/c = 2(L_{0i}+v_it)/c$ (i=1,2) where $L_{0i}$ represent the initial positions of the two targets. As such, the measured voltage at laser contacts at steady state can be shown to be proportional to the normalized carriers density variation [3, 18]:

$$\Delta N = \tau_p G_n (N - N_{th})$$
$$= -2\frac{\tau_p}{\tau_c}\left[k_1\cos(\omega_F\tau_1) + k_2\cos(\omega_F\tau_2)\right] \quad (2)$$
$$= -2\frac{\tau_p}{\tau_c}\left[k_1\cos(A_1 \pm \omega_1 t) + k_2\cos(A_2 \pm \omega_2 t)\right]$$

where $\tau_p$ is the photon lifetime, $G_n$ is the modal gain coefficient, $N_{th}$ is the threshold value of the carriers density, $A_i = 2L_{0i}\omega_F/c$, and $\omega_i = 2|v_i|\omega_F/c$, thus confirming that the measured signal contains information about speed (and consequently, position variation) of both targets.

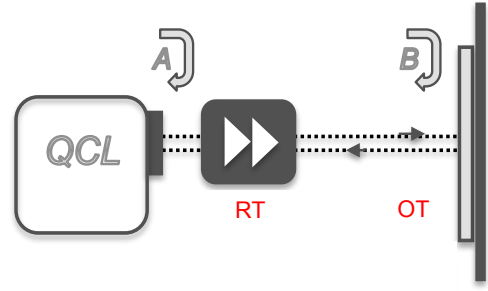

Fig. 1. Sketch of the common path sampling feedback interferometer.

While it may be objected that the use of stationary state solutions is improper for time-dependent delays, we remark that the slowest timescale on which the system evolves is of the order of few tens of nanoseconds [2, 19], so that the temporal signal of the carrier density, under translation of the two mirrors, adiabatically follows the stationary equations as long as the parameter changes are slower than the field and carrier relaxation timescales, i.e. when the fastest target moves with a speed smaller than 100 m/s. This has been verified by estimating numerically the transient relaxation time scale from dynamical integration of the LK equations.

As previously predicted and experimentally confirmed (see Figs. 2 and 4 in [17]), the power spectra of $\Delta N(t)$ in case of moderate and strong feedback show that a nonlinear frequency mixing is brought about by the implicit and transcendental character of Eq. (1), so that a spectral component can appear and even become dominant at the sum or the difference (for antiparallel or parallel translations, respectively) of the frequencies $\omega_1$ and $\omega_2$.

This nonlinear behavior is fundamental for the real-time measurement of OT displacement because it ensures that the temporal voltage signal contains a fast feature, linked nevertheless to the slow OT motion. We verified that this fingerprint is strictly due to the feedback strength provided by the combined targets by analyzing the power spectrum of the SM signal and specifically the line at the frequency difference $\omega_1-\omega_2$ for different values of feedback coefficients $k_1$ and $k_2$ by keeping fixed the ratio $k_1/k_2$ (this would correspond to add a neutral density filter in the common path A in Fig. 1).

The parameters used in our simulations and typical for a MIR-QCL are reported in Tab. 1.

TABLE I
TAB.1. PHYSICAL PARAMETERS USED IN THE LK MODEL

| $\tau_p$ | $\tau_c$ | $\alpha$ | $\omega_0$ | $L_{01}$ | $L_{02}$ |
|---|---|---|---|---|---|
| 100 ps | 35.6 ps | 2 | 302 THz | 0.15 mm | 0.25 mm |

Figure 2 shows the plot of the spectral power ratio Q between the amplitude of the peak at $\omega_1$ to that of the "nonlinear" line versus the strongest feedback $k_1$. While in the "linear" regime Q diverges (Q > 10 for $k_1 \leq 5\times10^{-3}$) because of the disappearance of the peak at the frequency difference, in the "nonlinear" regime we observe the existence of a power law linking Q and $k_1$ (see the inset in Fig. 2) due to the nonlinear coupling provided by the superposition of the laser field with the radiation reflected by the two targets. By increasing the feedback level the amplitude of the Fourier component at the frequency difference increases accordingly and it becomes dominant in the "strongly nonlinear" regime ($Q < 1$ for $k_1 > 2 \times 10^{-2}$). For $k_1 > 5 \times 10^{-2}$ the dependence of Q versus $k_1$ dependence deviates from a power law, possibly due to higher order effects.

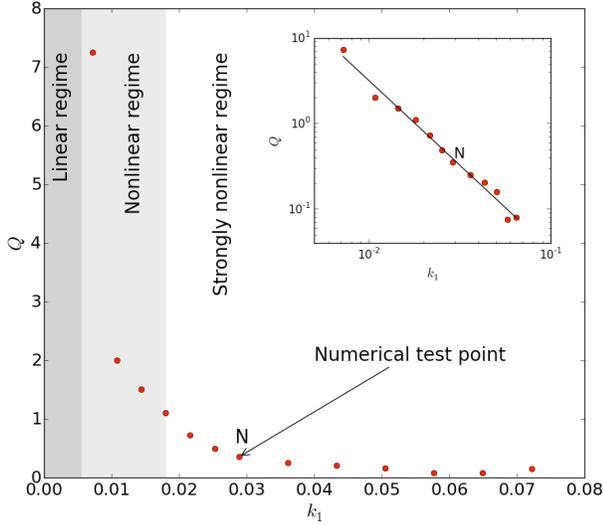
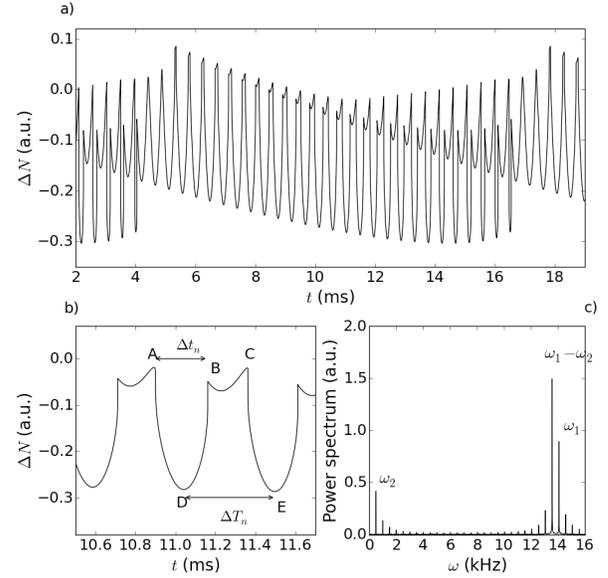

Fig. 2. Plot of the ratio Q between the amplitude of the frequency peak at $\omega_1$ and that of the frequency peak at $\omega_1$-$\omega_2$ versus the feedback coefficient $k_1$ ($k_1/k_2 = 1.18$). The two target velocities are $v_1 = 7$ mm/s and $v_2 = 0.25$ mm/s. Inset: log plot of Q vs $k_1$ in the "nonlinear" and "strongly nonlinear" regime. The solid line represents the best fit $Q \propto k_1^{-1.98}$.

Fig. 3. (a) SM signal $\Delta N$ vs time for $v_1 = 7$ mm/s, $v_2 = 0.25$ mm/s, $k_1 = 0.029$ and $k_2 = 0.025$ (point N in the strongly nonlinear regime of Fig. 2). (b) Zoom of the interferogram in Fig. 3a showing the presence of sub-features in the fast fringes. It is possible to show that the emergence of these sub-features can be related to the discontinuities of the emitted frequency $\omega_F$ with the variation of $L_1$ and $L_2$. The time interval between the minima D and E is denoted by the lower arrow and labeled $\Delta T_n$, represents the period of the fast fringe, and the time interval between A and B, denoted by the upper arrow and labeled $\Delta t_n$, is the time interval between two subsequent sub-features. (c) Power spectrum of $\Delta N$ with a main peak at the frequency difference $\omega_1$-$\omega_2$.

In principle, the spectral analysis of the SM signal in the nonlinear regime can thus provide information about OT speed, when RT is considered as a reference (and vice-versa), but clearly this requires data post-processing. To achieve real-time measurements, we must a) identify the signatures of OT motion in the time trace of $\Delta N(t)$ corresponding to the $\omega_1 \pm \omega_2$ component and b) relate these signatures to the OT displacement. In the following we derive such a relation showing that allows the estimation of the slow OT displacement with a resolution that depends on the fast RT displacement. In particular we demonstrate that by increasing the reference speed, we can achieve better resolution in the OT displacement up to 60 nm corresponding to about $\lambda/100$.

By numerically solving Eq. (1) for $\omega_F$, while adiabatically varying $L_1$ and $L_2$ at constant and parallel velocities, we obtain the time trace of the SM signal $\Delta N(t)$ shown in Fig. 3a with its power spectrum (Fig. 3c) for the parameters $v_1 = 7$ mm/s, $v_2 = 0.25$ mm/s, $k_1 = 0.029$, $k_2 = 0.025$, corresponding to point N in Fig. 2. The figure shows in detail that the presence of OT ($k_2 \neq 0$) induces the appearance of extra features (namely, the cusps pair B and C in Fig. 3b) beside the conventional sawtooth shape of the fast-switching part of the SM signal. In particular, we denote with $\Delta T_n$ and $\Delta t_n$ the temporal separation between two minima (labeled D and E in Fig. 3b) and two consecutive cusps (labeled A and B in Fig. 3b), of the n-th and n+1-th fast interference fringe respectively.

A first attempt to physically interpret these temporal features and to link them to the OT displacement can be performed by analyzing the LK steady state equations. In the hypothesis $v_2 \ll v_1$, we may assume that the minima of the fast fringes in $\Delta N$ occur at a time $T_n$ where the function $\cos(\omega \tau_1)$, appearing in the RHS of Eq. (2) has a maximum:

$$\omega_F \tau_1 = 2n\pi, n \in \mathbb{Z} \Rightarrow T_n \simeq \frac{1}{v_1}\left(\frac{n\pi}{\omega_F} - L_{01}\right)$$
$$= \frac{1}{v_1}\left(\frac{n\lambda_F}{2} - L_{01}\right) \quad (3)$$

where $\lambda_F$ represents the QCL wavelength in presence of feedback.

By using Eq. (1) we get:

$$\lambda_F = \frac{2\pi c}{\omega_0} + \frac{2\pi c}{\omega_0}\left(\frac{k_1}{\omega_0 \tau_c}[\alpha \cos(\omega_F \tau_1) + \sin(\omega_F \tau_1)]\right)$$
$$+ \frac{2\pi c}{\omega_0}\left(\frac{k_2}{\omega_0 \tau_c}[\alpha \cos(\omega_F \tau_2) + \sin(\omega_F \tau_2)]\right) \quad (4)$$
$$= \lambda_0 (1+\beta)$$

with:

$$\beta = \frac{k_1}{\omega_0 \tau_c}\left[\alpha \cos\left(\omega_F \frac{2L_1}{c}\right) + \sin\left(\omega_F \frac{2L_1}{c}\right)\right]$$
$$+ \frac{k_2}{\omega_0 \tau_c}\left[\alpha \cos\left(\omega_F \frac{2L_2}{c}\right) + \sin\left(\omega_F \frac{2L_2}{c}\right)\right] \quad (5)$$
$$= \beta(L_1) + \beta(L_2)$$

where $L_1$ and $L_2$ are the positions of RT and OT at time t. Then:

$$\Delta T_n \simeq (T_{n+1} - T_n) = \frac{1}{v_1}\left[(n+1)\frac{\lambda_F(T_{n+1})}{2} - n\frac{\lambda_F(T_n)}{2}\right]$$
$$= \frac{1}{v_1}\left[\frac{\lambda_0}{2} + (n+1)\frac{\lambda_0}{2}\beta(L_{2,n+1}) - n\frac{\lambda_0}{2}\beta(L_{2,n})\right]$$
$$\simeq \frac{1}{v_1}\left[\frac{\lambda_0}{2} + \delta_0 + \delta_1 \times (L_{2,n+1} - L_{2,n})\right] \quad (6)$$

where $L_{2,n+1}$ and $L_{2,n}$ represent the positions of OT corresponding to $T_{n+1}$ and $T_n$, respectively and we have formally introduced a first-order expansion of β with respect to the OT position $L_{2,n}$ in the hypothesis $v_1 \gg v_2$:

$$\beta(L_{2,n+1}) = \frac{k_2}{\omega_0 \tau_c}\left[\alpha \cos\left(\omega_F \frac{2L_{2,n+1}}{c}\right) + \sin\left(\omega_F \frac{2L_{2,n+1}}{c}\right)\right]$$
$$\simeq \beta(L_{2,n}) + \left(\frac{\partial \beta}{\partial L_2}\bigg|_{L_2 = L_{2,n}}\right) \times (L_{2,n+1} - L_{2,n}) \quad (7)$$

The parameters $\delta_0 = \frac{\lambda_0}{2}\beta(L_{2,n})$ and $\delta_1 = \frac{\lambda_0}{2}(n+1)\frac{\partial \beta}{\partial L_2}\bigg|_{L_2=L_{2,n}}$

that cannot be expressed in a closed analytical form, depend on $k_1$ and $k_2$, $\omega_F$ and $L_{2,n}$ through β and they may not be constant at subsequent instants of time. Yet, as confirmed by the numerical simulations, this can be safely assumed at this order of approximation. Equation (6) proves how the information about OT motion is encoded in the features of the fast fringes. For fixed $v_2$ Eq. (6) allows to recover the velocity $v_2$ by:

$$v_2 = \frac{1}{\delta_1}\left[v_1 - \frac{1}{\Delta T}\left(\frac{\lambda_0}{2} + \delta_0\right)\right] \quad (8)$$

where we omitted the dependence of ΔT on n because we verified that in this case these intervals remain constant throughout all the slow fringes to an excellent extent.

In Fig. 4 we represent the ratio $v_2/v_1$ versus the corresponding time interval between minima ΔT as estimated from different numerical simulations for $v_1$ =7 mm/s, $k_1$ = 0.029, $k_2$ = 0.025 (red dots). The figure reveals that the hyperbolic relation predicted by Eq. (8) and represented by the continuous line fits well the numerical results for $v_2$ > 30 μm/s ($v_2/v_1$ > 4.28×10$^{-3}$). The rms of the relative deviations between simulated and fitted data (denoted with ξ in the following) is of the order of 10$^{-4}$ and the corresponding fitted parameters are $\delta_0$ = −5.16×10$^{-9}$ m and $\delta_1$ = 0.57. Below the critical value of $v_2$ = 30 μm/s, indicated by the dashed line in Fig. 4, the OT speed seems to lose correlation to ΔT and accordingly ξ becomes larger that 1. As it will be clear in the following, the points corresponding to values of the ratio $v_2/v_1$ larger than 3×10$^{-2}$ are not represented in Fig. 4 because they violate the working hypothesis of $v_2 \ll v_1$.

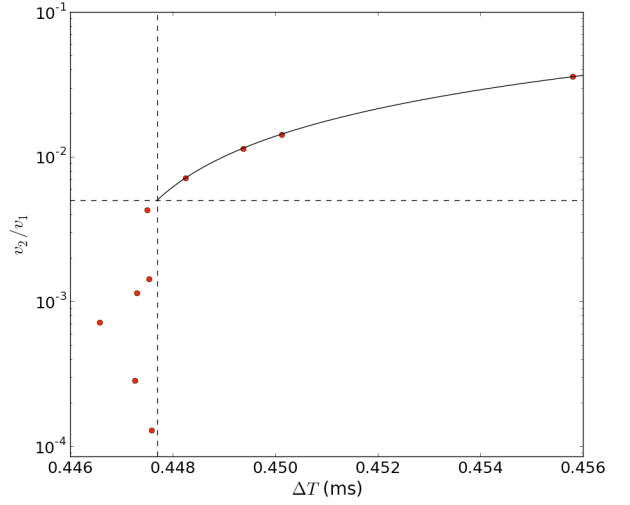

Fig. 4. Log-lin plot of the ratio $v_2/v_1$ *vs* ΔT. The solid line plots the hyperbolic relation predicted by Eq. (8), the red dots are the results of numerical simulations. The relation is satisfactorily verified down to $v_2$ = 30 um/s (dashed line). The other parameters are as in Fig. 3.

## III. NUMERICAL ANALYSIS

The above analysis shows that Eq. (8) offers a direct proof that OT translation can be reconstructed from fast features (the minima) in the time trace; yet it also fails for slow OT motion. This failure may be well ascribed to the approximations introduced in deriving Eq. (6).

In this section we will derive another relation linking the OT displacement *S* to temporal intervals in the time trace, without resorting to approximations of the steady state equations. To this purpose, let us stress that a formal dependence of S on the characteristics of the temporal trace described in Fig. 3 must exist, even if it is not known explicitly. Nevertheless, we can get a glimpse of this dependence by performing a fitting procedure, which will be illustrated in the following, valid for a wider range of $v_1$ and $v_2$ and, in principle, even for a generic OT motion.

To this purpose, we ran various simulations first keeping $v_1$ fixed at the value 7 mm/s and changing $v_2$ from 0.25 mm/s to 0.75 μm/s, and then keeping $v_2$ fixed at the value 0.3 mm/s while changing $v_1$ from 9 mm/s to 2.2 m/s. These intervals have been chosen both to provide a sufficient amount of data for the fitting procedure and to remain confidently distant from the limit of validity of the steady equations (1)-(2).

For what concerns the temporal feature, to which the displacement will be linked in this procedure, we considered the cusp pairs that emerge in the fast fringes (see points B and C in Fig. 3b) which are both a notable legacy of the nonlinear mixing of the feedback from the two targets (Fig. 3c) and easy to track by an algorithm the selects the extrema. All simulations show that, starting from the beginning of a slow fringe, the temporal interval $\Delta t_n$ between two cusps of consecutive fringes (as indicated in Fig. 3b) decreases as the OT moves and eventually vanish at the end of the slow fringe when, depending on the feedback value, few fast fringes could miss the sub-features (see Fig. 3a). We will discuss this issue later in the section. This behavior clearly indicates that $\Delta t_n$ depends on the relative motion of the two targets and thus it represents a valid candidate to measure $v_2$ when $v_1$ is known.

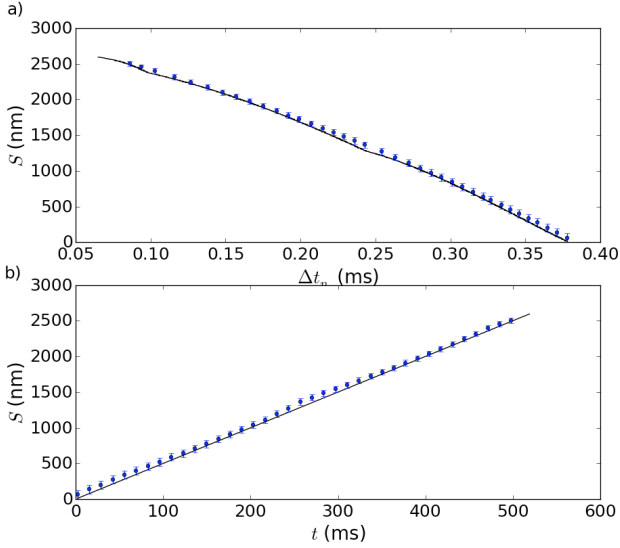

Fig. 5. Comparison between the theoretical displacement of the OT, represented by the solid lines, and the values obtained from Eq. (9) plotted with the corresponding error bars as a function of $\Delta t_n$ (a) and t (b). Velocities set for the RT and the OT are, respectively, $v_1 = 7$ mm/s and $v_2 = 5$ μm/s (only one point every 30 is drawn). The other parameters are as in Fig. 3.

Since the OT velocity is constant, the theoretical position of the target grows in time according to $S_{th}=v_2 t_B$ where $t_B$ are the times at which the leftmost cusp of the sub-feature is recorded, taking as t = 0 the first recorded cusp in the slow fringe. For the moment being, we will focus on only one slow fringe (on the extension to more slow fringes will be discussed later in the section).

The aim of the procedure is to obtain a predicted OT position $S = f(v_1, \Delta t_n)$. Among various tested fitting functions, the best fit of the simulation sets (at fixed $v_1$) was proved to be a quadratic polynomial of the form $S = C_2 \Delta t_n^2 + C_1 \Delta t_n + C_0$ with values of $C_i$ that do not change significantly for different $v_2$. Then, in order to make explicit the coefficient dependence on $v_1$, we performed a similar fitting procedure on the sets of simulations with constant $v_2$. The final form expressing the OT displacement S as a function of cusp intervals, for arbitrary $v_1$ reads then:

$$S = -\gamma_2 \frac{v_1^2}{\lambda_0}\Delta t_n^2 - \gamma_1 v_1 \Delta t_n + \gamma_0 \frac{\lambda_0}{2} \quad (9)$$

where $\gamma_2 = 1.06 \pm 0.03$, $\gamma_1 = 0.64 \pm 0.01$, $\gamma_0 = 0.94 \pm 0.01$.

The parameter values are obtained as the average over all the values derived from the various simulations, and the errors are their standard deviations. The rms of the relative deviations between simulated and fitted data is of order $10^{-2}$ for any $v_2 > 2$ μm/s, while below this threshold it rapidly grows, thus indicating that the fitting procedure can no longer be considered reliable.

The procedure described above allows us to measure the OT displacement as long as the algorithm used to identify the sub-features can assign a definite value to the time interval $\Delta t_n$. Yet, as we mentioned above, just at the end of each slow fringe there can occur a few fast fringes that miss the sub-feature. This "blinds" the acquisition process for a short time lapse. In order to overcome this limitation, we can extrapolate the value of the displacement starting from the last tracked time where the sub-feature is detectable over the "blind" interval until the reappearance of the sub-features. The extrapolation is made using the simple formula $\Delta S = v_{last} \Delta t_{blind}$, where $v_{last}$ is the OT velocity as evaluated by the algorithm up to the start of the "blind" time interval $\Delta t_{blind}$. This also allows us to follow the OT motion across the edge of the single slow fringe and connect the procedure to the following one, thus addressing the other issue mentioned earlier. In section IV we will show an implementation of this procedure on several slow fringes from actual experimental data, which gives results well within the desired accuracy.

An example of application of our approach is given in Fig. 5. Panels (a) and (b) show the position of the OT, given by the formula (9) with the corresponding error bars, as a function of $\Delta t_n$ and t respectively, for $v_1 = 7$ mm/s and $v_2 = 5$ μm/s. Note that the OT velocity is an order of magnitude below the limit found for the reliability of Eq. (8). The solid lines represent the theoretical displacements $S_{th}$ derived from the ideal law $S_{th} = v_2 t$.

It also follows from Eq. (9) that, in order to estimate the OT displacement, we need to identify at least two consecutive cusps B, so that the minimum measurement time is of the order of the fast fringe period $\lambda_0/2v_1$, giving a corresponding minimum measurable displacement

$$\Delta S_{min} \simeq \frac{v_2}{v_1}\lambda_0 \quad (10)$$

As an example we report in Fig. 6a and Fig. 6b two close-ups of S(t) corresponding to the OT velocity $v_2 = 5$ μm/s and two RT velocities $v_1 = 50$ μm/s and $v_1 = 500$ μm/s, respectively. We note that while the smallest measurable displacement in the configuration of Fig. 6a is about 250 nm, the increased number of points in Fig. 6b allows for OT displacement resolution of around 60 nm ($\lambda = 6$ μm).

The procedure described so far clearly shows that the fast moving RT acts like a "stroboscope" with adjustable period for the measurement of the OT position. In principle the resolution of this technique increases with the ratio $v_2/v_1$ within the constraints $v_1 < 100$ m/s and $v_2 > 2$ μm/s derived above. However, the largest error source for this method is the uncertainty in the values of the parameters $\gamma_i$, which is of order $10^{-2}$, the effective accuracy is limited to around $\lambda/100$.

We finally observe that a realistic sensor that exploits this technique to measure sub-wavelength displacement should be first calibrated to find correct values for the $\gamma_i$ parameters, as it is often the case for relative, as opposed to absolute, sensors. This can be done by comparing the measure of the displacement with the actual position of an OT moving with known fixed speed. The procedure needs to be done only once, and our result shows that the parameters obtained in this way can be used for all the OT and RT velocities compliant with the physical limits discussed above.

In the following section we report on an experimental validation of Eq. (9).

## IV. EXPERIMENTAL RESULTS

### A. Experimental setup. Self-mixing configuration

The schematic of the homodyne SM configuration sensing is shown in Fig. 1. Basically, it consists of a single-arm multiple-cavity interferometer. Phase and amplitude modulations of the reinjected field, induced by changes in the two target positions, alter the laser operations, producing a voltage drop ($V_{QCL}$) across the QCL active region.

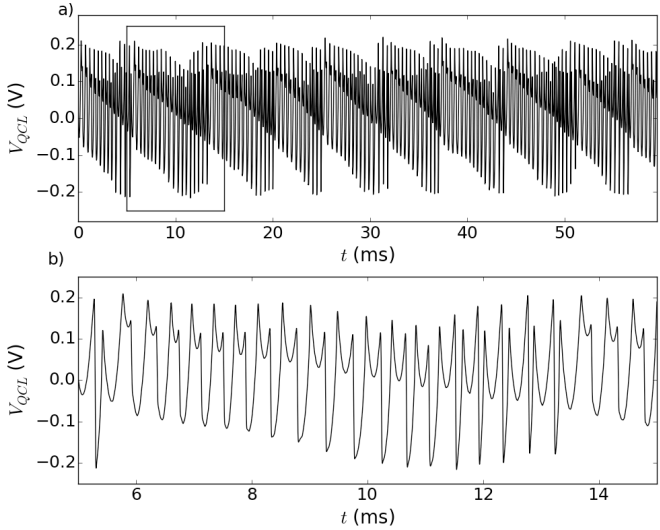

Fig. 7. (a) Representative oscilloscope trace of the voltage modulation detected at the QCL terminals (scope time base setting: 200 ms/div; sample rate: 2.5 MS/s). The SM waveform was obtained in response to forward translation of both oscillator and target ($v_1 = 5$ mm/s and $v_2 = 0.25$ mm/s). (b) Close up of the SM signal in panel (a) (highlighted region) showing the presence of the fringe sub-features.

The quantum cascade laser was a single longitudinal mode emitting at $\lambda_{QCL} \approx 6.2$ μm and temperature stabilized at 283 K. The sensitivity to optical feedback was optimized by driving the QCL slightly above the threshold at constant current I = 490 mA (i.e. $I_{th} = 486.5$ mA for the solitary QCL). The output beam was collimated by an AR-coated chalcogenide glass aspheric lens having numerical aperture NA = 0.56 and nominal focal length of 4 mm. The laser source was coupled to a OT and a RT inserted in the optical path, as shown in Fig. 1. In details, a trial external cavity configuration was arranged, with feedback contributions coming simultaneously from the optical paths A, B respectively. A partially transparent polypropylene thin sheets (83% transmittance at 6.2 μm) were used as RT and OT for the specific experimental arrangement in Fig. 1. Also, the laser beam was conveniently attenuated by inserting foils of polymethylpentene along the optical axis (not shown in the figure) to adjust the effective strength of optical feedback. The voltage offset measured across the device (i.e. the incoherent amount of back-scattered radiation) was subtracted by ac-coupling to a voltage amplifier with gain 40 dB.

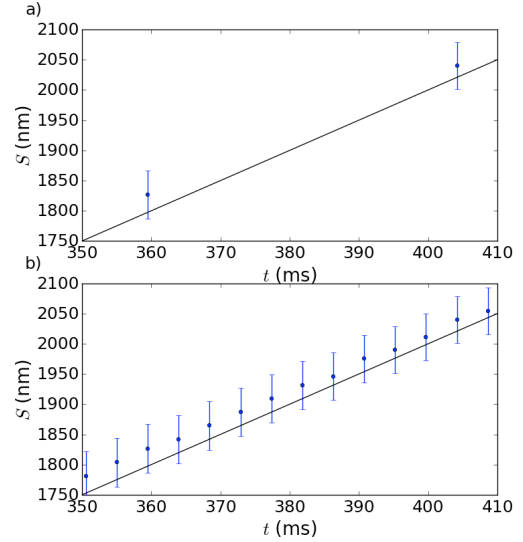

Fig. 6. Focus of S vs t in Fig. 5b for the same OT velocity and two different RT velocities: $v_1 = 50$ μm/s for the plot (a) and $v_1 = 500$ μm/s for the plot (b).

### B. Results

Figure 7 shows a representative SM interferometric waveform $V_{QCL}$ produced by translating the RT along the optical axis with a stepper-motor stage (NEWPORT M-VP-25XA), and simultaneously moving the OT at a lower velocity in the same direction. The set velocity was $v_1= 5$ mm/s and $v_2 = 0.25$ mm/s, respectively. Comparable features hold regardless of the motion direction.

The laser attenuation was changed until the sub-features became clearly visible in the time trace, and gated for coupling conditions, which maximized the fringe contrast.

In Fig. 7b, we expand the waveform in Fig. 7a, to appreciate the complex interference pattern and the fringe sub-features.

We note a very good qualitative agreement with the SM signal resulting from the simulations (compare with Fig. 3).

Figure 8 shows the estimation of the OT displacement S versus t (colored symbols) as given by the application of the phenomenological formula (9) to the experimental data in Fig. 7. The four plots correspond to four evaluations of S, where the calibration stage for estimating the $\gamma_i$ was performed on 2,3 and 4 consecutive slow fringes, respectively (out of the nine available, see Fig. 7a). We observe that the maximum deviation between the reconstructed OT motion and the OT motion with constant speed set with the translation stage (solid line) is of few hundreds of nanometers and thus it is widely compliant with the mechanical fluctuations and deviations from linear translations as specified by the manufacturer of the stage.

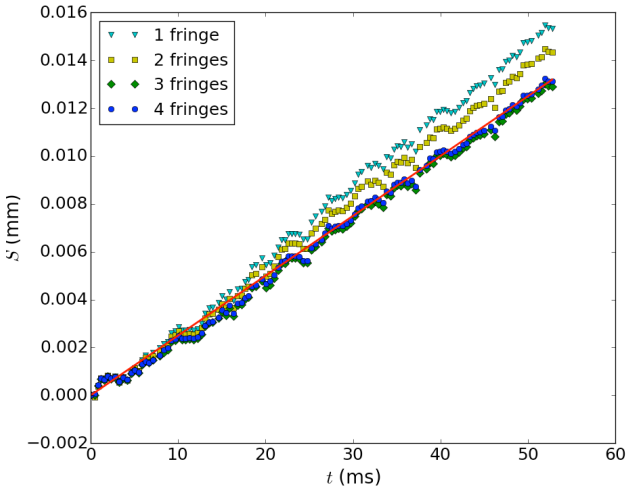

Fig. 8. Displacement S of the OT as derived by applying the formula (9) to the experimental data in Fig. 7 using interpolation on 1, 2, 3, 4 OT fringes (colored symbols). The red line indicates the imposed OT translation $S_{exp} = v_2 t$.

## V. CONCLUSION AND POSSIBLE DEVELOPEMENTS

In conclusion, we showed that the SM signal from a QCL with feedback can exhibit nonlinear mixing of the two frequencies associated with the speeds of two independently moving targets. The strong re-injection needed for the mixing regime maintains the laser in a stable emission state due to the enhanced stability of QCLs. This phenomenon makes the fast fringes in the temporal behavior of the SM signal to exhibit features depending not only on the higher speed (associated with the RT) but also on the lower one (associated with the OT). We demonstrated that an approximated analytical expression for such dependence can be derived from the LK model describing the system and provides a fitting algorithm to measure the OT displacement in real-time with a non-optimized accuracy around $\lambda/100$. The experimental results confirm that the OT displacement can be measured from the fast waveform features and the above procedure can be reliably implemented to give a measure compatible with the translator stage accuracy.

As possible developments of this technique, we remark two issues that we are presently considering. The first one concerns the possibility of increasing the feedback ratio and thus increment the 'strength' of the nonlinear mixing (see Fig. 2 and Fig. 3b), in terms of the spectral component of frequency combinations (possibly of higher order such as $2\omega_1-\omega_2$, etc.). Such spectral features imply the occurrence of other sub-features in the temporal trace [20], other than the one characterized in our paper, e.g. in Fig. 3. The time separation of such sub-sub-features reduces further the 'stroboscope' interval allowing for denser measurements and pushing the resolution to even lower limits. Of course this will require the formulation of an entirely new algorithm for an interpolating formula equivalent to Eq. (9), taking into account (e.g.) two time intervals now, the former sub-feature $\Delta t_n$ and the new sub-sub-feature periodicity.

The second issue concerns the possibility to extend this technique to RT motions with bounded spatial extent, such as harmonic oscillations or the like (and, in perspective, to arbitrary target dynamics). Such extension requires to study in detail the interferometric time trace from such a configuration (oscillating RT and translating OT), correlating it to the spectral signatures of the nonlinear frequency mixing and eventually selecting the significant intervals to be extracted from the time trace and to be interpolated with an obviously completely different function, other than Eq. (9). Whether the more complicated RT dynamics, involving accelerations, may hinder the accuracy and whether the latter can be restored by more accurate or different calibration stages, is still to be assessed.

## VI. ACKNOWLEDGMENTS

The authors acknowledge financial support from MIUR – PON02-0576 INNOVHEAD and MASSIME, PON01-02238 EURO6, COST Action BM1205 and the Fondazione Caripuglia, Research project "Studio di sorgenti laser QCL per la realizzazione di sensori avanzati"

The authors also wish to thank G. Ficco for contributing to the data analysis.

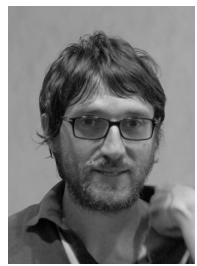

**Francesco P. Mezzapesa** received the M.S. degree in physics (2001) at the University of Bari, Italy, where he is currently Research Associate since 2010. His research activities include THz QCL-based sensing and imaging; fundamental physics of QCLs; laser self-mixing interferometric sensors for mechatronic system diagnostics; ultrafast laser micro- and nano-processing for smart sensing. In 2007, he was awarded the degree of PhD at the Optoelectronics Research Centre of the University of Southampton (UK), where his main research domains were integrated optics and nonlinear photonic microstructures. From 2006 to 2008, he worked as Industrial Project Manager at Sintesi Scpa (Bari, Italy), mainly on the development of innovative sensing solutions for applications in optomechatronics. Before joining the Physics Department in Bari, he had a two-year PostDoc position in the Soft Matter Nanotechnology group at the CNR-Nano in Lecce

**Giuseppe De Risi** was born in Bari, Italy, in 1977. He received his M.S. degree in physics from University of Bari in 2000 and his Ph.D. in physics from University of Perugia in 2003. He was granted postdoctoral fellowships in University of Bari, at CERN and at the ICG, University of Portsmouth. He is currently a postdoc at the Polytechnic of Bari.

Until 2012 his research interests deals with string and brane cosmology, then, after a postgraduate course in computer science, he turns to the study of the dynamics of QCL lasers with feedback.

**Lorenzo L. Columbo** was born in Bari in 1975. He received the M.S. degree in physics in 2002 at the University of Bari (IT) and the Ph.D degree in Physics in 2007 at the University of Insubria, Como (IT). From 2005 since 2006 Lorenzo Columbo visited the Computational Nonlinear and Quantum Optics group at the University of Strathclyde, Glasgow (GB). After a two years post-doc at the Institut Nonlineaire de Nice (FR) he won a research grant at the National Research Council (CNR) in Bari. In 2011 Lorenzo Columbo got a three years research contract at the University of Insubria, Como where he was one of the principal investigators of a national project for young scientists founded by the Italian Minister for University and Research. From May 2014 to date he is associate researcher at the Department of Physics of the University of Bari. He is also associate with the Institute for Photonics and Nanotechnology (IFN).

His theoretical research activity focuses on pattern formation and structure localization in semiconductors lasers and nonlinear optical resonators for applications mainly in the field of all-optical information storage and processing. Since 2009 he also works on modeling self-mixing interferometry in conventional semiconductor lasers and Quantum Cascade lasers for sensing and imaging applications. Lorenzo Columbo carries on several national and international collaborations with some of the most active experimental and theoretical groups in the field of nonlinear optics in dissipative extended systems (Institut Nonlineaire de Nice, University of Como, University of Rome "Sapienza" (IT), University of l'Aquila (IT), etc..).

He is the author of more than 30 refereed papers.

**Massimo Brambilla** was born in Milan, Italy in 1961. He received the Laurea in Physics at the University of Milan (IT) in 1988 and the Dokt.Phil. at the University of Zuerich (CH) in 1992. From 1988 to 1995 he has earned several grants and postdoc positions with the National Institute for the Physics of Matter (INFM), the National Research Council (CNR) and has been visiting scientist at various Universities and research centers in Europe. From 1995 to 2001 he was Assistant Professor at the Polytechnic of Bari and from 2001 to date he is Associate Professor in Matter Physics at the same University.

He is the author of more than 100 refereed journals and over

60 proceedings. His research interests include the spatiotemporal dynamics of nonlinear optical systems, laser and semiconductor laser physics, optical information treatment and optical solitons.

Massimo Brambilla is a member of the Optical Society of America and referee for over twelve main journals in the field of Physics and optics.

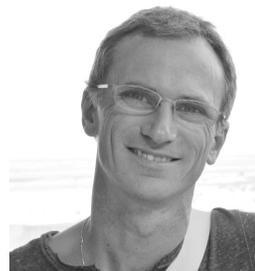

**Maurizio Dabbicco** received his Physics ('89) and Ph.D. ('93) degrees from the University of Bari, Italy. For two years ('94-'95) he was Research Assistant at the Clarendon Laboratory, University of Oxford, UK. He joined the Physics Department of the University of Bari in '96 and became Professor of Experimental Physics in '05.

His research interests embrace semiconductor nonlinear and quantum optics, development of new optical characterization techniques for optoelectronic materials, novel applications of photonic and optoelectronic devices in the field of mechatronics and industrial metrology. He has co-authored about 80 papers and four patent applications. He is currently referee of Applied Optics, Optics Letters, Applied Physics B, IEEE Sensors and Photonics Technology Letters, and project evaluator of the Italian Ministry of Research and University (MIUR).

His professional activity include: organizer and editor of the International Workshop on Coherent Effects on Elementary Excitations in Semiconductors ('97), coordinator and chief investigator of several national and international projects also in partnership with leading industrial companies.

M. Dabbicco was awarded a National Research Council (CNR) Fellowship (1992) and an European Union Training Fellowship (1994). He is an Associate staff member of the Institute of Photonics and Nanotechnologies of CNR and of some national and international optics societies.

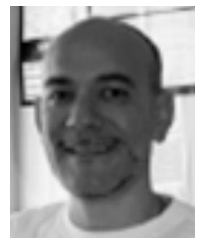

**Vincenzo Spagnolo** received the Ph.D in physics, in 1994 from University of Bari. From 1997 to 1999 he worked as researcher of the National Institute of the Physics of Matter (INFM). From 1999 to 2003, he was a Postdoctoral Research Associate at the Physics Department, University of Bari. Since January 2004 he works as assistant Professor of Physics at the Polytechnic of Bari.

His research interests include quantum cascade lasers, spectroscopic techniques for real-time device monitoring, optoacoustic gas sensors.

His research activity is documented by more than 130 publications and 2 filed patents. He has given more than 30 invited presentations at international conferences and workshops.

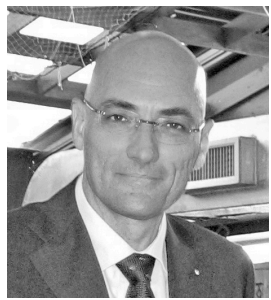

**Gaetano Scamarcio** received the PhD in physics from the University of Bari, Italy, in 1989.

Since 2002 he is full professor of experimental physics at the University of Bari, Italy. From 1989 to 1990 he has been a research fellow at the Max-Planck-Institute für Festkörper-forschung, Stuttgart, Germany, and in 1992 a visiting scientist at the Walter-Schottky-Institute, Garching, Germany. In the period 1994-1996, in 2000 and 2001 he has been a visiting scientist of Bell Laboratories, Lucent Technologies (formerly AT&T), Murray Hill, NJ (U. S. A.). In 2006 he has been invited professor at the University of Paris 7.

His research interests include the development and applications of quantum cascade lasers, optical, vibrational and transport properties of semiconductor structures at the nanoscale, spectroscopic techniques for real-time monitoring of optoelectronic devices, optoelectronic sensors for mechatronics.

Gaetano Scamarcio was the recipient of the Award of the Italian Physical Society in 1989, the Firestone Prize for young laureates in 1985, a NATO-CNR Advanced Fellowship in 1995, and the Ambassadorial Fellowship of the International Rotary Foundation in 1994.